\documentclass[%
reprint,
superscriptaddress,
 amsmath,amssymb,
 aps,
prc,
]{revtex4-1}

\usepackage{graphicx}% Include figure files
\usepackage{dcolumn}% Align table columns on decimal point
\usepackage{bm}% bold math
\usepackage{color}
\usepackage{url}
\def\be{\begin{eqnarray}}   
\def\ee{\end{eqnarray}}

\newcommand{\cvec}[1]{\reflectbox{\ensuremath{\vec{\reflectbox{\ensuremath{#1}}}}}}

\newcommand{\affA}{
Max Planck Institute for the Structure and Dynamics of Matter and Center for Free-Electron Laser  Science, Luruper Chaussee 149, 22761 Hamburg, Germany
}
\newcommand{\affB}{
Department of Chemistry, Dalhousie University, 6274 Coburg Road Halifax, Nova Scotia B3H 4R2, Canada
}
\newcommand{\affC}{
Center for Computational Quantum Physics (CCQ), The Flatiron Institute, 162 Fifth avenue, New York NY 10010
}
\newcommand{\affD}{
Nano-Bio Spectroscopy Group, Universidad del Pa\'is Vasco, 20018 San Sebasti\'an, Spain
}

\begin{document}

\preprint{APS/123-QED}

\title{Coupled forward-backward trajectory approach
for non-equilibrium electron-ion dynamics}% Force line breaks with \\

\author{Shunsuke A. Sato}\affiliation{\affA}
\author{Aaron Kelly}\affiliation{\affB}
\author{Angel Rubio}\affiliation{\affA}\affiliation{\affC}\affiliation{\affD}

\date{\today}% It is always \today, today,

\begin{abstract}

We introduce a simple ansatz for the wavefunction of a many-body system based on coupled forward and backward-propagating semiclassical trajectories. This method is primarily aimed at, but not limited to, treating nonequilibrium dynamics in electron-phonon systems. The time-evolution of the system is obtained from the Euler-Lagrange variational principle, and we show that this ansatz yields Ehrenfest mean field theory in the limit that the forward and backward trajectories are orthogonal, and in the limit that they coalesce. We investigate accuracy and performance of this method by simulating electronic relaxation in the spin-boson model and the Holstein model. Although this method involves only pairs of semiclassical trajectories, it shows a substantial improvement over mean field theory, capturing quantum coherence of nuclear dynamics as well as electron-nuclear correlations. This improvement is particularly evident in nonadiabatic systems, where the accuracy of this coupled trajectory method extends well beyond the perturbative electron-phonon coupling regime. This approach thus provides an attractive route forward to the \textit{ab-initio} description of relaxation processes, such as thermalization, in condensed phase systems.

\end{abstract}

\maketitle

%\tableofcontents

%=========================================================================================
\section{Introduction \label{sec:intro}}
%=========================================================================================

The nonequilibrium dynamics of an interacting system of nuclei and electrons is one of the most fundamental subjects in condensed matter physics, and it plays an central role in many applications of current interest such as photosynthesis \cite{Nature.446.782,Nature.463.644}, photovoltaics \cite{Nature.414.338,Nature.comm.7.11755}, proton transfer \cite{bell2013proton}, light-induced phase transitions \cite{Nature.449.72,Nature.530461}, laser processing \cite{Science.279.208,malinauskas2016ultrafast}, and many more. Developing an accurate and numerically efficient description of the nonequilibrium dynamics of coupled electron-nuclear systems is essential to understand the microscopic mechanisms underlying such phenomena. However, despite of the significance of the potential applications, theoretical descriptions are severely limited due to the computational expense associated with the simulation of realistic systems.

One possible route forward is to adopt a mixed quantum-classical approach where the electronic dynamics 
is treated quantum mechanically while the nuclear dynamics is treated on a semi-classical level. These approaches can be rigorously justified on the basis of the large disparity between nuclear and electronic masses, for example. Indeed, the Ehrenfest mean field (MF) dynamics method \cite{McLachlan64}, which is one of the simplest of the mixed quantum-classical approaches, has been combined with \emph{ab-initio} electron dynamics simulations based on the time-dependent density functional theory (TDDFT) and applied to investigate quite large systems thus far \cite{PhysRevLett.97.126104,Science.344.1001,Nature.comm.4.1602}. The \emph{ab-initio} nonadiabatic molecular dynamics is another successful \emph{ab-initio} approach
based on TDDFT and a surface hopping algorithm \cite{J.Chem.Phys.93.1061,PhysRevLett.98.023001} and has been applied to various phenomena \cite{curchod2013trajectory,C3CP51514A}

However, the MF approach does not retain predictive power in many cases of interest \cite{WHM2009,Grunwald2009,mfgqme}, primarily due to the neglect of correlations in the dynamics. As a result, a hierarchy of trajectory-based dynamics approaches have been developed over the past decades which attempt to improve on the accuracy of mean-field theory. These approaches include methods based on propagating the density matrix, such as the quantum-classical Liouville equation\cite{qcle99} and associated approximations such as the Poisson bracket mapping equation \cite{PBME}, and the forward-backward trajectory solution \cite{JChemPhys138.134110,JChemPhys137.22A507}, as well as the closely related linearized and partially linearized path integral approaches \cite{WHM2001,LSCIVR,Land-map,PLDM,ILDM}. In addition, methods based on wavefunction propagation, such as the coupled coherent states method and the multi-configurational Ehrenfest approaches \cite{CCS, MAKHOV2017200} of Shalashilin and co-workers, the multi-Davydov ansatz methods \cite{JChemPhys.137.084113,JChemPhys.143.014113}, and coupled trajectory quantum-classical approaches \cite{PhysRevLett.115.073001} based on an exact factorization of the electron-nuclear wavefunction have also been developed. Although these sophisticated methods have succeeded in surpassing the accuracy of mean field theory in model systems, applications to realistic systems based on \emph{ab-initio} treatments have been severely limited by computational cost. In order to achieve a comprehensive \emph{ab-initio} description of nonequilibrium electron-nuclear dynamics in realistic systems, accuracy, efficiency, and simplicity must be balanced.

In this work, we propose a simple ansatz for the many-body wavefunction of an electron-nuclear system, in order to capture the quantum nature of the dynamics beyond the mean field level. Our prescription is based on the variational principle, and it involves pairs of coupled semi-classical trajectories; one of which is propagating forward in time, and the other backward. The proposed method is one of the simplest possible extensions of the Ehrenfest mean field dynamics method, which utilizes only individual trajectories to construct ensemble averages. 

We examine the performance of our new approach by treating nonequilibrium electronic relaxation processes in two paradigmatic model systems: the single-mode spin-boson model \cite{Leggett87,PhysRev.49.324,PhysRev.51.652,PhysRevLett.107.100401}, and the Holstein model \cite{PhysRevB.91.104302,HOLSTEIN1959325}. The single-mode spin-boson model is one of the simplest nontrivial models for a coupled quantum system, and is also known as the Rabi model, or the Jaynes-Cummings model. Despite its apparent simplicity, it captures a range of rich phenomena \cite{JCM} and has been intensively investigated in various contexts such as in quantum optics \cite{JCM,PhysRevLett.108.163601}, and superconductivity \cite{PhysRevLett.96.127006,Nature.445.515}. The Holstein model is the modern work-horse model for the description of electron phonon coupling effects in solids, such as polaron formation and transport \cite{PhysRevA.84.051401,PhysRevLett.110.223002}, and photo-carrier relaxation \cite{PhysRevB.91.104302}. Despite of the apparent simplicity of our proposed method, we find that it substantially improves upon the performance of the Ehrenfest dynamics method, and that it accurately captures the quantum nature of the nuclear dynamics in many of the cases studied. Thus, the concept of coupling forward and backward propagating semi-classical trajectories could be a key tool in developing an accurate and efficient theoretical treatment for use in applications to realistic systems and hence, this method could be used as a base to extend the accuracy of \textit{ab-initio} Ehrenfest dynamics simulations for future practical applications.

The construction of remainder of this paper is as follows: In Sec. \ref{sec:method}, 
we introduce our wavefunction ansatz and develop the associated evolution equations from the Euler-Lagrange variational principle. In Sec. \ref{sec:results} we examine nonequilibrium electronic relaxation dynamics within the single-mode spin-boson model and the Holstein model. We compare our proposed method with the multi-trajectory Ehrenfest dynamics method, the forward-backward trajectory solution to the quantum-classical Liouville equation, as well as numerically exact quantum mechanical results. Finally, our findings are summarized in Sec. \ref{sec:summary}.

%=========================================================================================
\section{Theory \label{sec:method}}
%=========================================================================================

In this section, we introduce an ansatz for the wavefunction that is based on the multi-trajectory Ehrenfest dynamics method. We then derive equations of motion using the Euler-Lagrange variational principle. We call this approach the \textit{coupled forward-backward trajectory (CFBT)} method for reasons that will become clear below.

First, we consider a general quantum subsystem coupled to an external environment (bath). The total system is described by the following Hamiltonian:
\be
\hat H = \hat H_s + \hat H_b + \hat H_{sb},
\ee
where $\hat H_s$ and $\hat H_b$ describe the Hamiltonian of the subsystem and the bath,
respectively. The coupling between the subsystem and the bath is described 
by $\hat H_{sb}$.

For notational convenience, let us assume that the bath Hamiltonian can be decomposed into a harmonic part and a residual part $\Delta \hat w$ as follows:
\be
\hat H_b = \sum^{N_{b}}_{n=1} \frac{\hbar \omega_n}{2} \left (
\hat a^{\dagger}_n  \hat a_n + \frac{1}{2}
\right )
+ \Delta \hat w.
\ee
Using the annihilation operator $\hat a_n$, 
a coherent state can be defined as 
$\hat a_n |z_n \rangle = z_n |z_n \rangle$.
For convenience, we introduce the following notation 
for direct products of coherent states;
$|z\rangle = |z_1\rangle \otimes \cdots \otimes |z_{N_b}\rangle$,
where $z$ is a generalized coordinate; $z:= \left \{z_1, \cdots, z_{N_b} \right \}$.

Now, consider the time-evolution of an arbitrary observable, $\hat B(t)$:
\be
\left \langle \hat B(t) \right \rangle = 
\mathrm{Tr} \left [
\hat B(t) \hat \rho
\right ],
\ee
where $\hat \rho$ is the density matrix of the entire system.
Inserting the closure relations for the subsystem space and  
the coherent states, the observable can be described by,
\be
\left \langle \hat B(t) \right \rangle &=& 
\sum_{\alpha \beta}
\int \frac{d^2z}{\pi^{N_b}} \int \frac{d^2z'}{\pi^{N_b}}
\left \{ \langle \beta | \otimes \langle z'|  \right \}
\hat \rho
\left \{  |\alpha \rangle \otimes |z \rangle \right \}
\nonumber \\
&& \times
\left \{ \langle \alpha | \otimes \langle z|  \right \}
\hat B(t)
\left \{  |\beta \rangle \otimes |z' \rangle \right \}.
\label{eq:td-observable}
\ee

Here, we focus on the integrand of Eq. (\ref{eq:td-observable}),
\be
&& \left \{ \langle \alpha | \otimes \langle z|  \right \}
\hat B(t)
\left \{  |\beta \rangle \otimes |z' \rangle \right \} \nonumber \\ 
&&= \left \{ \langle \alpha | \otimes \langle z|  \right \}
\hat U^{\dagger}(0,t) \hat B \hat U(0,t)
\left \{  |\beta \rangle \otimes |z' \rangle \right \},
\label{eq:integrand}
\ee
where the forward propagator, $\hat U(0,t)$,
and the backward propagator, $\hat U^{\dagger}(0,t)=\hat U(t,0)$,
are involved.
In order to construct an approximation for Eq. (\ref{eq:integrand}),
we introduce a linear combination of the forward and backward 
propagated wave functions with a phase factor,
\be
|\psi (t,\theta) \rangle
= \hat U(0,t) | \alpha \rangle \otimes | z \rangle  
+ e^{i \theta} 
\hat U(0,t) | \beta \rangle \otimes |  z' \rangle.   \nonumber \\
\label{eq:wf-phase-sum}
\ee

One can prove that the phase average of the expectation value for $|\psi (t,\theta)\rangle$
is reduced to Eq. (\ref{eq:integrand}):
\be
&&\frac{1}{2\pi} \int^{2\pi}_0 d \theta e^{-i\theta}
\langle \psi(t,\theta) | \hat B | \psi(t,\theta) \rangle  \nonumber \\
&&= \left \{ \langle \alpha | \otimes \langle z|  \right \}
\hat B(t)
\left \{  |\beta \rangle \otimes |z' \rangle \right \}.
\ee
Therefore, the observable of Eq. (\ref{eq:td-observable}) 
can be rewritten in the following phase-averaging form:
\be
\left \langle \hat B(t) \right \rangle &=& 
\sum_{\alpha \beta}
\int \frac{d^2z}{\pi^{N_b}} \int \frac{d^2z'}{\pi^{N_b}}
\left \{ \langle \alpha | \otimes \langle z|  \right \}
\hat \rho
\left \{  |\beta \rangle \otimes |z' \rangle \right \}
\nonumber \\
&& \times
\frac{1}{2\pi} \int^{2\pi}_0 d \theta e^{-i\theta}
\langle \psi(t,\theta) | \hat B | \psi(t,\theta) \rangle.
\label{eq:td-observable-phase-average}
\ee

Since no approximations have employed up to this point, Eq. (\ref{eq:td-observable-phase-average}) is a formally exact expression.
For practical applications, however, one needs to approximate the propagator $\hat U(0,t)$. For this purpose, we approximate the time-propagation of the wave function $|\psi(t,\theta)\rangle $ by assuming the following simple ansatz,
\be
|\tilde \psi(t)\rangle = |\alpha(t)\rangle \otimes | z(t) \rangle
+ |\beta(t) \rangle \otimes | z'(t) \rangle,
\label{eq:ansatz-wf}
\ee
where the total wavefunction $|\tilde \psi(t)\rangle$ is expressed by 
a sum of two factorized wave functions.
Note that the phase factor $e^{i \theta}$ in Eq. (\ref{eq:wf-phase-sum})
is absorbed in the degree of freedom of $|\beta(t)\rangle$ of the ansatz,
and the phase contribution is taken into account via initial conditions of $|\beta(t)\rangle$.

The equation of motion for the ansatz wave function can be derived from 
the following Lagrangian,
\be
L = i\hbar \frac{
\langle \tilde \psi(t)| \dot {\tilde \psi} (t)\rangle
-\langle \dot {\tilde \psi}(t)| \tilde \psi (t)\rangle 
}{2}
- \langle \tilde \psi(t)| \hat H | \tilde \psi (t)\rangle. 
\label{eq:lagrangian00}
\nonumber \\
\ee

One can derive the equation of motion for the subsystem state $|\alpha(t)\rangle$ 
by the Euler-Lagrange equation,
\be
\frac{d}{dt}\frac{\partial L}{\partial \langle \dot \alpha(t)|}
-\frac{\partial L}{\partial \langle \alpha(t)|} = 0.
\label{eq:euler-lagrange-sub}
\ee
The derived equation of motion is
\be
&& i\hbar|\dot \alpha(t)\rangle + i\hbar|\dot \beta(t)\rangle \langle z(t)| z'(t)\rangle
\nonumber \\
&+& i\hbar| \beta(t)\rangle \langle z(t)| \dot z'(t)\rangle 
-\hbar|\alpha (t) \rangle \Im \left [ \langle z(t)| \dot z (t)\rangle  \right ] \nonumber \\
&=& \hat H_{eff}(z,z) |\alpha(t) \rangle + \hat H_{eff}(z,z') |\beta(t) \rangle,
\label{eq:eom-sub}
\ee
where the effective Hamiltonian is $\hat H_{eff}(z,z') = \langle z| \hat H | z'\rangle$.
Detailed derivation of Eq. (\ref{eq:eom-sub}) is 
described in Appendix \ref{appendix:deriveation}.
Similar equations can be derived for $|\beta(t)\rangle$, 
and one can construct a matrix form for the equations of motion,
%\be
\be
i \hbar S_b \frac{d}{dt} 
\left(
    \begin{array}{c}
      |\alpha(t)\rangle   \\
      |\beta(t)\rangle
    \end{array}
  \right)
= \left [
H_{eff} - \hbar D_b 
\right ]
\left(
    \begin{array}{c}
      |\alpha(t)\rangle   \\
      |\beta(t)\rangle
    \end{array}
  \right),
\label{eq:eom-sub-mat}
\ee
where $S_b$, $D_b$, and $H_{eff}$ are the following $2\times2$ matrices:
\be
S_b = \left(
    \begin{array}{cc}
      1 & \langle z | z' \rangle  \\
      \langle z' | z \rangle & 1
    \end{array}
  \right),
\ee
\be
D_b = \left(
    \begin{array}{cc}
      \Re \left[ i \langle z|\dot z \rangle \right] & i \langle z | \dot z' \rangle  \\
      i\langle z' | \dot z \rangle & \Re \left[i \langle z'|\dot z' \rangle \right] 
    \end{array}
  \right),
\ee
and
\be
H_{eff} = \left(
    \begin{array}{cc}
      \hat H_{eff}(z,z) & \hat H_{eff}(z,z')  \\
      \hat H_{eff}(z',z) & \hat H_{eff}(z',z')
    \end{array}
  \right).
\ee
Note that the norm conservation in Eq. (\ref{eq:td-observable-phase-average})
is guaranteed by Eq. (\ref{eq:eom-sub-mat}).

One can also derive the equation of motion for the coherent states from,
\be
\frac{d}{dt}\frac{\partial L}{\partial  \dot z^*_n} 
- \frac{\partial L}{\partial z^*_n} = 0,
\label{eq:euler-lagrange-bath}
\ee
and hence;
\be
\frac{\partial}{\partial z^*_n}\langle \tilde \psi | \hat H |
\tilde \psi \rangle &=& i\hbar \dot z_n \langle \alpha|\alpha\rangle 
+ i\hbar z_n \Re\left[ \langle \alpha| \dot \alpha \rangle \right] \nonumber \\
&-& \hbar z_n \Re\left [ 
i \langle \alpha | \dot \beta \rangle \langle z|z'\rangle
+i \langle \alpha | \beta \rangle \langle z| \dot z'\rangle
\right ] \nonumber \\
&+& i\hbar \dot z'_n\langle \alpha| \beta \rangle \langle z | z'\rangle
\nonumber \\ &+& i\hbar z'_n \left \{
\langle \alpha | \dot \beta \rangle \langle z | z'\rangle
+\langle \alpha | \beta \rangle \langle z | \dot z'\rangle
\right \}. 
\label{eq:eom-bath}
\ee
The detailed derivation is described in Appendix \ref{appendix:deriveation}.

Here, we note the close relationship between the CFBT method and Ehrenfest dynamics. In the limit that the coherent states are orthogonal, $\langle z(t)|z'(t)\rangle = 0$, the equations of motion of the CFBT method reduce to the equations of motion for Ehrenfest dynamics. On the other hand, in the perfect overlap limit where $\langle z(t)|z'(t)\rangle = 1$, the forward and backward trajectories coalesce, and Eqs. (\ref{eq:euler-lagrange-sub}) and (\ref{eq:euler-lagrange-bath}) again yield Ehrenfest mean field dynamics. Indeed, one can derive the evolution equations for Ehrenfest mean field theory from Eq. (\ref{eq:td-observable}), by assuming (i) the initial density matrix is not entangled, (ii) the orthogonal relation for the coherent states, $|\langle z | z'\rangle|^2 \approx \pi^{N_b} \delta(z-z')$, and (iii) the single-trajectory wave function ansatz, $|\tilde \psi(t) = |\alpha (t)\rangle \otimes | z(t)\rangle$, where the wavefunction is described by the direct product of a subsystem state $|\alpha(t)\rangle$ and the bath coherent state $|z(t)\rangle$. A detailed derivation of these two results is provided in Appendix \ref{appendix:mtef}. As the CFBT method does not assume any orthogonality condition for the coherent states, and instead employs a generalized ansatz, it can be seen as a generalization of Ehrenfest dynamics where the coupling between trajectories allows for deviations from mean field behavior.

For harmonic baths,  $\Delta w = 0$, furthermore, in the case of bilinear system-bath coupling (as in the electron-phonon problems studied here), the coupling part of the Hamiltonian has a sum-of-products structure, \\ $\hat H_{sb} = -\gamma \sum_n \left (\hat a^{\dagger}_n +\hat a_n \right) 
\otimes \hat \Gamma_{n}$, where $\hat{\Gamma}_n$ are linear operators that act only on the subsystem. In this case, the left hand side of Eq. (\ref{eq:eom-bath}) can be rewritten as,
\be
\frac{\partial}{\partial z^*_n}\langle \tilde \psi | \hat H |
\tilde \psi \rangle &=& \hbar \omega_n z_n \langle \alpha| \alpha \rangle 
+ \hbar \omega_n z'_n
\langle \alpha | \beta \rangle \langle z| z'\rangle \nonumber \\
&-&z_n \Re \left [\langle \alpha, z | \hat H 
|\beta , z' \rangle \right ]
+ z'_n  \langle \alpha, z |\hat H | \beta , z' \rangle
\nonumber \\
&-& \gamma \Big(  \Gamma_{n, \alpha \alpha} 
+ \Gamma_{n,\alpha \beta} \langle z | z'\rangle\Big).
\label{eq:force_for_cs}
\ee
The detailed derivation of Eq. \ref{eq:force_for_cs} is described 
in Appendix \ref{appendix:deriveation}.

Combining Eq. (\ref{eq:eom-bath}), Eq. (\ref{eq:force_for_cs}), and
similar expressions for $z'_n$, one can obtain
the following matrix expression for the equation of motion 
in the case of a harmonic bath with bilinear coupling;
\be
i\hbar S \frac{d}{dt}
\left(
    \begin{array}{c}
      z_n(t)   \\
      z_n'(t)
    \end{array}
  \right)
&&= \left [ \hbar \omega_n S + E - \hbar D \right ]
\left(
    \begin{array}{c}
      z_n(t)   \\
      z_n'(t)
    \end{array}
  \right)  \nonumber \\
&&- \gamma \left(
    \begin{array}{c}
      \Gamma_{n, \alpha \alpha} 
      +\Gamma_{n, \alpha \beta}  \langle z|z'\rangle   \\
        \Gamma_{n, \beta \beta}   
      + \Gamma_{n, \beta \alpha} \langle z'|z\rangle  
    \end{array}
  \right), \nonumber \\
\label{eq:eom-bath-mat}
\ee
where $S$, $E$, and $D$ are the following $2\times2$ matrices,
\be
S = \left(
    \begin{array}{cc}
      \langle \alpha | \alpha \rangle 
& \langle \alpha | \beta \rangle \langle z | z' \rangle  \\
      \langle \beta | \alpha \rangle\langle z' | z \rangle 
& \langle \beta | \beta \rangle 
    \end{array}
  \right),
\ee
\be
E = \left(
    \begin{array}{cc}
      -\Re \left [\langle \alpha,  z  |\hat H | \beta, z' \rangle \right ]
& \langle \alpha,  z  |\hat H | \beta, z' \rangle  \\
      \langle \beta,  z'  |\hat H | \alpha, z \rangle
&       -\Re \left [ \langle \beta,  z'  |\hat H | \alpha, z \rangle \right ]
    \end{array}
  \right), \nonumber \\
\ee
and
\be
D &=& \left(
    \begin{array}{cc}
      i \Re \left [ \langle \alpha | \dot \alpha \rangle \right  ] & 0  \\
      0 & i \Re \left [ \langle \beta | \dot \beta \rangle \right  ]
    \end{array}
  \right) \nonumber \\
&+& 
\left(
    \begin{array}{cc}
      - \Re \left [ i\langle \alpha, z | \vec{\frac{d}{dt}} |\beta,z'  \rangle \right  ] 
& i\langle \alpha, z | \vec{\frac{d}{dt}} |\beta,z'  \rangle  \\
      i\langle \beta, z' | \vec{\frac{d}{dt}} |\alpha,z  \rangle
& - \Re \left [ i\langle \beta, z' | \vec{\frac{d}{dt}} |\alpha,z  \rangle \right  ]
    \end{array}
  \right), \nonumber \\
\ee
where $\vec{\frac{d}{dt}}|\alpha,z\rangle$ denotes 
$\vec{\frac{d}{dt}}|\alpha,z\rangle=|\dot \alpha \rangle \otimes | z\rangle + |\alpha\rangle \otimes |\dot z\rangle$.

By self-consistently solving Eq. (\ref{eq:eom-sub}) and Eq. (\ref{eq:eom-bath}), or equivalently 
Eq. (\ref{eq:eom-sub-mat}) and Eq. (\ref{eq:eom-bath-mat}) for bilinear-harmonic problems, one can propagate the ansatz wavefunction of Eq. (\ref{eq:ansatz-wf}). 
The construction of time-dependent observables, Eq. (\ref{eq:td-observable-phase-average}), can then be easily carried out.

In order to evaluate the integrals of Eq. (\ref{eq:td-observable-phase-average})
for the phase $\theta$ and the phase spaces $\{z,z'\}$, we employ a basic Monte Carlo
sampling procedure. For the sampling of the  phase, $\theta$ , we generate a pair of phases, $\theta = \phi$ and $\phi+\pi$, where $\phi$ is drawn from
a uniform random distribution between $0$ and $2\pi$. For the $\{z,z'\}$ phase space sampling, 
we sample from the following correlated Gaussian distribution,
\be
G_c(z,z') = \left (\frac{\sqrt{3}}{2\pi} \right )^{2N_b}
\exp \left [
-\frac{|z|^2 + |z'|^2  + |z-z'|^2}{2}
\right ], \nonumber \\
\label{eq:correlated-gauss}
\ee
where $|z|^2$ denotes $\sum_n |z_n|^2$.

This correlated Gaussian distribution (\ref{eq:correlated-gauss}) is related to the integrand of Eq. (\ref{eq:td-observable}) for a pure subsystem operator, which contains the inner product of two 
coherent states; $\langle z|z'\rangle = \exp \left [ -|z-z'|^2/2 -i \Im [zz'^*] \right ]$,
where $zz'^*$ denotes $\sum_n z_nz'^*_n$.
However, 
as this overlap integral also contains a complex phase factor, the Monte Carlo sampling procedure based on sampling directly from Eq. (\ref{eq:correlated-gauss}) requires a larger number of trajectories to converge  than the MTEF method. In order to overcome this inefficiency in sampling for further applications, more sophisticated sampling methods need to be developed.

%=========================================================================================
\section{Results \label{sec:results}}
%=========================================================================================

In this section we examine the performance of the CFBT method, derived above in Sec. \ref{sec:method}, in treating the nonequilibrium electronic dynamics of the spin-boson model and the Holstein model.
We will compare the results of the CFBT method with those of the multi-trajectory Ehrenfest dynamics (MTEF) method, the forward-backward trajectory solution (FBTS) \cite{JChemPhys137.22A507,JChemPhys138.134110}, as well as the exact solution. 
In the MTEF method, which is one of the simplest mixed quantum-classical approaches,
dynamics of an observable is evaluated by 
ensemble average of Ehrenfest trajectories. In each trajectory, 
a subsystem is treated quantum mechanically with the Schr\"odinger equation, 
while a bath is treated fully classically with the Newton equation, 
as described in Appendix \ref{appendix:mtef}.
In the FBTS method, dynamics of an observable is also evaluated by 
ensemble average of semi-classical trajectories. However, these trajectories are different
from Ehrenfest trajectories, but their dynamics is based on an approximation to the formal solution to the quantum classical Liouville equation. Theoretical and numerical details of the FBTS
method are described elsewhere 
\cite{JChemPhys137.22A507,JChemPhys138.134110,martinez2015assessment}.
In both the MTEF and the FBTS methods, the ensemble average of trajectories
can be performed by Monte Carlo sampling for the initial condition of each trajectory.
Because we will only consider harmonic baths in this paper, the Monte Carlo
sampling in the both methods will be simply performed by the Gaussian distribution.
For the exact solution of the spin-boson model, we directly solve the time-dependent Schr\"odinger equation. For the Holstein model, benchmark data generated with the limited functional space method is taken from the reference \cite{PhysRevB.91.104302}. For simplicity, $\hbar$ will be taken as being equal to $1$ hereafter.

%=========================================================================================
\subsection{Single-mode spin-boson model}
%=========================================================================================

The Hamiltonian of the single-mode spin-boson model can be written as
\be
\hat H = \frac{\Delta}{2}\hat \sigma_x + \omega_0 \hat a^{\dagger}\hat a
+ \gamma \hat \sigma_z \otimes \left ( \hat a^{\dagger} + \hat a \right ),
\ee
where $\hat \sigma_x$ and $\hat \sigma_z$ are Pauli spin matrices.
In the single-mode spin-boson model, a two-level quantum subsystem with energy gap of $\Delta$ is coupled to a single harmonic oscillator with frequency $\omega_0$, via a bilinear coupling with strength $\gamma$.
Here, we set the initial state of the subsystem to be the direct product of the up-spin diabatic state and the ground state of the harmonic oscillator.
In this work, we consider the nonequilibrium electronic dynamics in a resonant regime, $\omega_0/\Delta = 1$.

\begin{figure}[htbp]
\centering
\includegraphics[width=0.8\columnwidth]{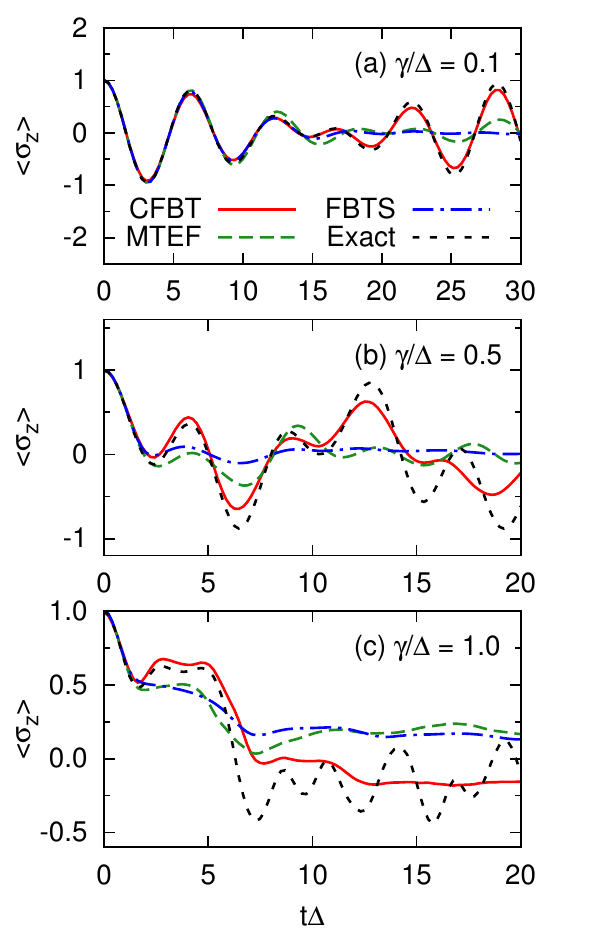}
\caption{\label{fig:Ekin_spin_boson}
Population dynamics of the single-mode spin-boson model 
in the resonant regime, $\omega_0/\Delta = 1$, with varying coupling strength $\gamma/\Delta$ from weak (a) to strong (c).
CFBT (red line), MTEF (green dashed line), FBTS (blue dash-dotted line), and the numerically exact solution (black dotted line).}
\end{figure}

Figure \ref{fig:Ekin_spin_boson} shows the population dynamics, $\langle \hat \sigma_z (t) \rangle$, for different coupling strength $\gamma/\Delta$. Panel (a) shows the result for $\gamma/\Delta = 0.1$, which corresponds to a weakly coupling regime. One can see that all the three approximated methods reproduce the oscillation and damping behavior in the exact result up to $t\Delta = 15$ fairly well. After $t\Delta = 15$, one can see the recurrence of the population dynamics in the exact result. While the MTEF and FBTS methods fail to describe the recurrent dynamics, the CFBT method reproduces this behavior exactly. This fact indicates that the CFBT method correctly captures the quantum coherence of bath dynamics.

In the panels (b) and (c), the results for stronger coupling are shown. Although the CFBT method shows deviation from the exact solution, still it provides the most accurate results among the three approximated methods. Especially for the strong coupling regime $\gamma/\Delta = 1.0$ in the panel (c), one can see that the CFBT method reproduces the exact result up to $t\Delta = 5$ extremely well, while the MTEF and FBTS failed to accurately capture the qualitative structure. Furthermore, the population dynamics of the CFBT method follows the mean value
of the oscillating exact population quite well for the long time region.
These facts indicate that the CFBT method substantially captures 
subsystem-bath correlation and significantly improves the dynamics 
even in the strong coupling regime.

%=========================================================================================
\subsection{One-dimensional Holstein model}
%=========================================================================================

We further examine the performance of the CFBT method in the context of the one-dimensional Holstein model. The Hamiltonian of the Holstein model is
\be
\hat H = \hat H_{kin} + \hat H_{ph} + \hat H_{coup},
\ee
where $\hat H_{kin}$ is the electronic kinetic energy, $\hat H_{ph}$
is the phonon energy, and $\hat H_{coup}$ is the electron-phonon
coupling. These terms are explicitly given by
\be
\hat H_{kin} &=& - t_0 \sum_j \left (
c^{\dagger}_j c_{j+1} + c^{\dagger}_{j+1} c_j
\right ), \\
\hat H_{ph} &=&  \omega_0 \sum_j a^{\dagger}_j a_j, \\
\hat H_{coup} &=& -\gamma \sum_j \left (
a_j + a^{\dagger}
\right ) \hat n_j,
\ee
where $\hat n_j$ is the electron number operator at $j$th-site;
$\hat n_j = c^{\dagger}_jc_j$.
In this model, the electron-phonon coupling strength is usually characterized by the following dimensionless parameter \cite{PhysRevB.91.104302}:
\be
\lambda = \frac{\gamma^2}{2t_0\omega_0}.
\ee 

We examine the relaxation dynamics of a 12-site chain with periodic boundary conditions. We first focus on an intermediate parameter regime, with $\lambda = 0.2$. This is a non-perturbative electron-phonon coupling regime, and hence rather difficult to capture accurately using approximated methods \cite{PhysRevB.91.104302}. We set the initial condition to be an uncorrelated product of the highest excited state of the electronic Hamiltonian, $\hat H_{kin}$, with the ground state of the phononic Hamiltonian, $\hat H_{ph}$, at zero temperature.

\begin{figure}[htbp]
\centering
\includegraphics[width=0.8\columnwidth]{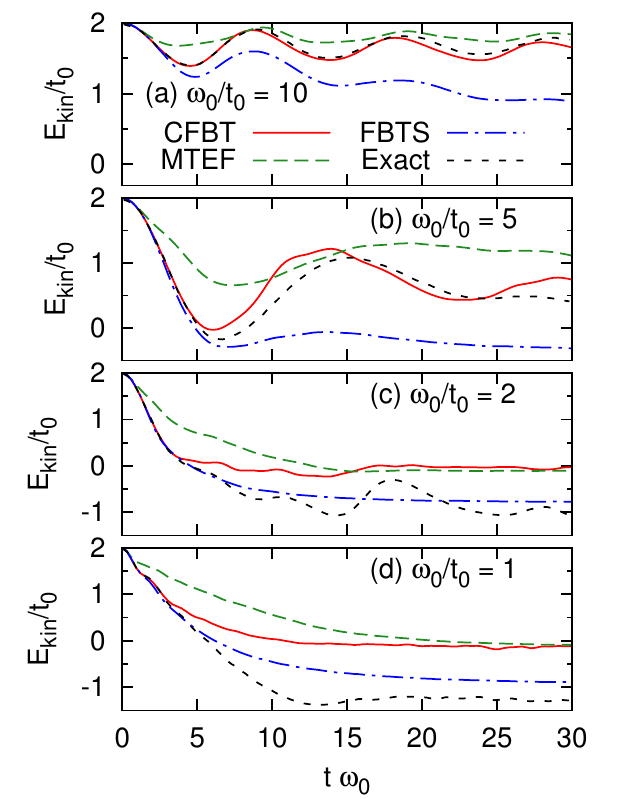}
\caption{\label{fig:Ekin_holstein}
Electronic kinetic energy dynamics in the Holstein model, for systems with varying nonadiabaticity ($\omega_0/t_0$) from strong, (a), to weak, (d). The numerically exact solution (exact) is taken from Ref. \cite{PhysRevB.91.104302}. Line styles are as given in Fig. 1.}
\end{figure}

Figure \ref{fig:Ekin_holstein} shows the electronic kinetic energy dynamics, 
$E_{kin}(t) = \left \langle \hat H_{kin}(t) \right \rangle$, for different nonadiabacity ratios $\omega_0/t_0$. Panel (a) shows the result for $\omega_0/t_0=10$, which is a strongly nonadiabatic regime with $\omega_0/t_0 \gg 1$. One can see that the CFBT method nicely reproduces the exact solution in this case, while the other methods fail to capture the correct behavior. As the quantum nature of bath is expected to play a significant role in the nonadiabatic regime, the performance of the CFBT method in Fig. \ref{fig:Ekin_holstein} (a) is surprisingly accurate, which indicates the importance of the coupling between the forward-backward trajectory pairs in capturing the quantum nature of the phonon dynamics. 

In the panels (b)-(d), the results for smaller nonadiabaticity ratios are shown. One sees that the CFBT method shows deviations from the exact solution for these cases, that can be particularly pronounced at long times. This onset of inaccuracy can be explained by the accumulation of the electron-phonon correlation during the relaxation process. In the adiabatic regime, where $\omega_0 \ll 4t_0$, numerous electron-phonon collisions must occur during the nonequilibrium dynamics in order for the small energy quanta of the phonon bath to accommodate the relaxing electronic system. However, as each electron-phonon collision induces some correlation between these degrees of freedom, a large degree of correlation can manifest. Therefore, such electronic relaxation processes in the adiabatic regime are challenging to describe with techniques that treat electron-phonon correlations approximately. 

\begin{figure}[htbp]
\centering
\includegraphics[width=0.8\columnwidth]{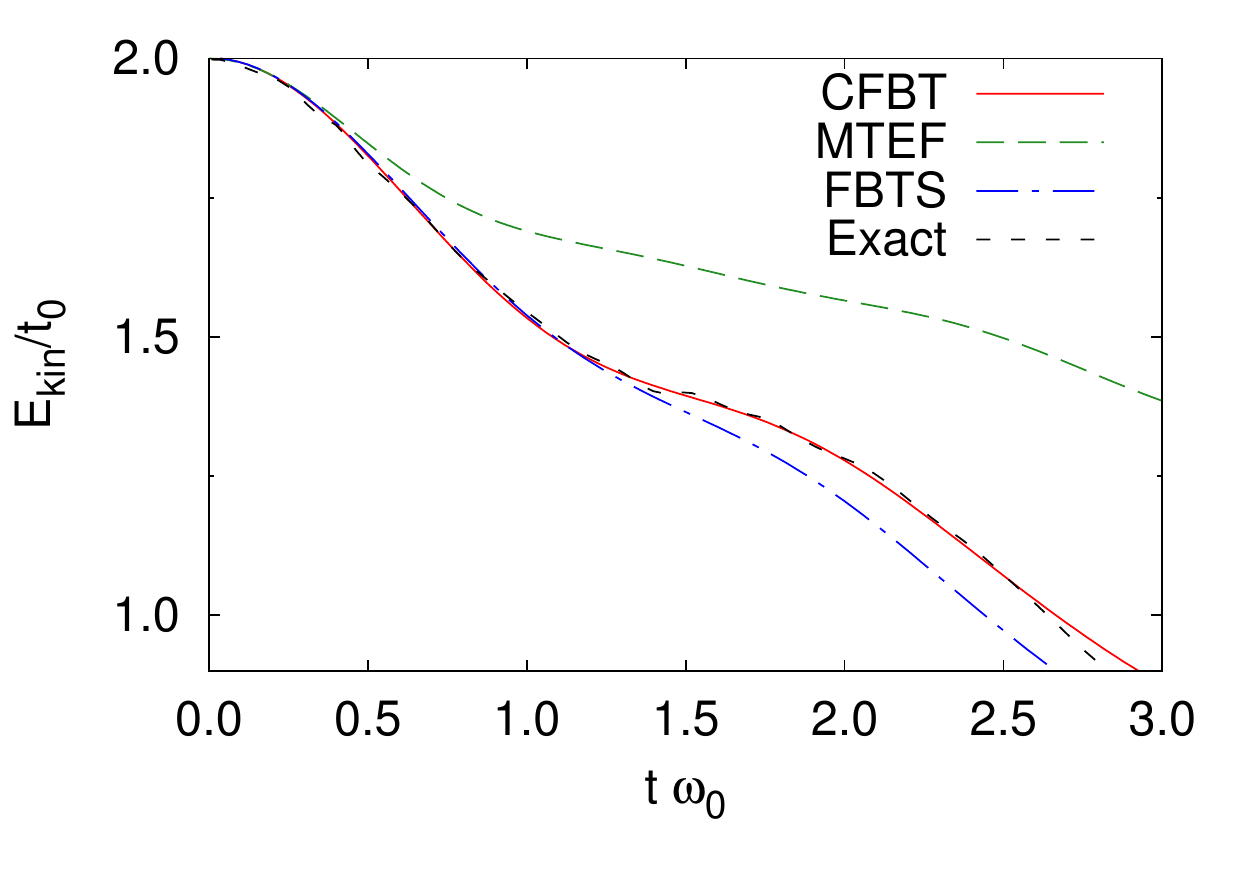}
\caption{\label{fig:Ekin_holstein_short}
Short time kinetic energy dynamics of the Holstein model.}
\end{figure}

Figure \ref{fig:Ekin_holstein_short} shows the short-time electronic relaxation dynamics in the Holstein model in the adiabatic regime, $\omega_0/t_0 = 1$, enlarged from Fig. \ref{fig:Ekin_holstein} (d). Although the CFBT method fails to describe the long-time dynamics in this regime due to the correlations that manifest, it does show the best short-time behavior among the three approximate methods we studied. While the FBTS and MTEF methods start to deviate from the exact result around $t\cdot \omega_0 = 1.5$ and $0.5$ respectively, the CFBT follows the exact result up to around $t\cdot \omega_0 = 2.5$. This indicates that the CFBT method captures some important non-adiabatic aspects of the electron-phonon correlation stemming from low-order scattering processes.

The generalized quantum master equation (GQME) opens an alternate path to achieve an accurate description of highly correlated nonequilibrium dynamics, by casting the effect
of the bath in terms the memory kernel \cite{PTP.20.948,JChemPhys33.1338}.
Both approximate and exact quantum dynamics methods have been employed to construct 
the memory kernel subsequently obtain the system dynamics from the GQME \cite{shi-geva,CohenRabani,mjgqme,CohenReichmanRabani,ThossRabani,mfgqme2,andresI,andresII}. In many condensed phase problems, the memory kernel has been shown to decay rapidly compared with the time scale of the associated relaxation dynamics. Hence, in practice one only needs the short time component of the memory kernel in order to compute the full nonequilibrium dynamics. Indeed, Kelly \textit{et al} have shown the memory kernel can orders of magnitude shorter-lived than the electronic population relaxation dynamics in the spin-boson model across the adiabatic and nonadiabatic regimes \cite{mjgqme,mfgqme}. This indicates that the highly accurate short-time CFBT data shown here could potentially be used to construct the memory kernel and generate the long-time dynamics via the GQME approach. 

As indicated from Fig. \ref{fig:Ekin_holstein} (a), the CFBT method can be highly accurate 
in a strongly nonadiabatic limit $t_0/\omega_0 \ll 1$. To elucidate this fact, we investigate
the electron dynamics in a strongly-coupled strongly-nonadiabatic regime
with $t_0/\omega_0=0.001$ and $\gamma/\omega_0=1$, corresponding to $\lambda=500$.
In the highly nonadiabatic limit, Dorfner \textit{et al} , derived 
an analytical expression for the electronic kinetic energy as a function of time
based on a perturbation theory \cite{PhysRevB.91.104302}. In the present case,
the kinetic energy is expressed as
\be
E_{kin}(t) = 2t_0e^{2g^2\left [ \cos(\omega_0 t) -1\right ]},
\label{eq:Ekin-antiadiabatic-analytic}
\ee
where $g$ denotes $\gamma/\omega_0$. Furthermore, they have demonstrated that
the analytical expression of Eq. (\ref{eq:Ekin-antiadiabatic-analytic})
is quantitatively accurate for the present parameter set ($t_0/\omega_0=0.001$ and $\gamma/\omega_0=1$).

\begin{figure}[htbp]
\centering
\includegraphics[width=0.8\columnwidth]{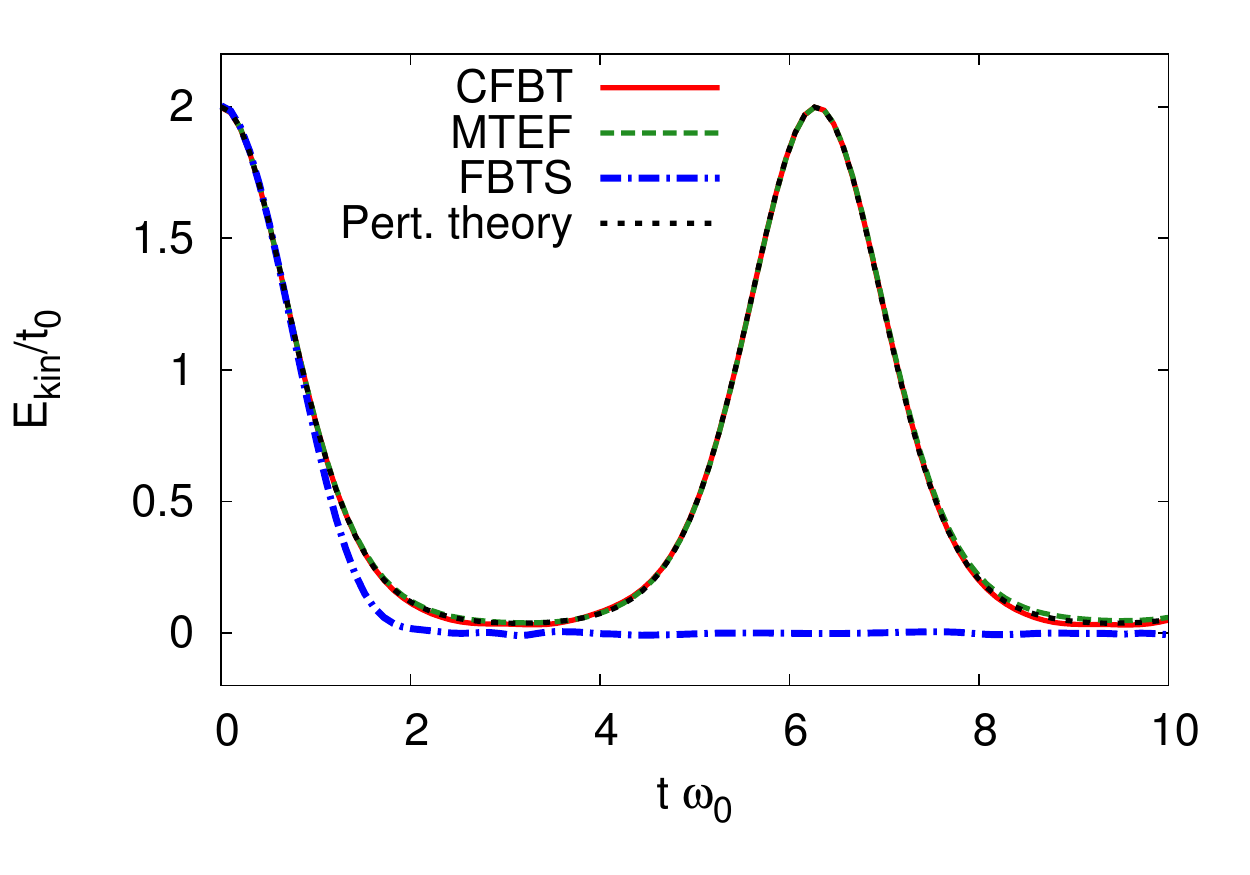}
\caption{\label{fig:Ekin_holstein_antiadiabatic}
The electronic kinetic energy dynamics of the Holstein model 
in the strong-coupling highly-nonadiabatic regime
with $t_0/\omega_0=0.001$ and $\gamma/\omega_0=1$, corresponding to $\lambda=500$.}
\end{figure}

Figure \ref{fig:Ekin_holstein_antiadiabatic} shows the electronic kinetic
energy dynamics $E_{kin}(t)$ with different methods. As expected,
one can confirm that the CFBT method accurately reproduces the exact result.
Furthermore, the MTEF method also accurately reproduces the exact result,
while the FBTS method fails to reproduce it even qualitatively.
The accurate description of the CFBT and MTEF methods can be explained by
less electron-phonon collision processes and hence less electron-phonon correlation
in the highly nonadiabatic regime. The large discrepancy between the energy quanta of the electronic and phononic systems, strongly suppresses energy exchange via scattering processes. Thus, the electronic subsystem and the phonon bath remain largely uncorrelated,
and the total system can be well described by a direct product state. As explained above, and in Appendix 
\ref{appendix:mtef}, the classical trajectories of the MTEF method
rely on the direct product wavefunction ansatz $|\alpha\rangle \otimes |z\rangle$.
Therefore, quantum direct product states may be well described in the MTEF as well as the CFBT method by construction. 

On the other hand, the failure of the FBTS method in this case is more likely due to the significant quantum nature of bath at zero temperature in this highly nonadiabatic regime. Although the FBTS method is also based on a rigorously derivable semi-classical propagator that generates similar equations of motion to the MTEF method, one of the approximations to the full quantum propagator which leads to FBTS trajectories does not correctly capture the zero-point motion (ground state) of the bath degrees of freedom. Kelly \textit{et al} have observed a similar breakdown of the FBTS method in the zero temperature spin-boson model with an Ohmic environment, while the MTEF method again retains a much higher degree of accuracy \cite{mfgqme}. This is also consistent with the performance of the FBTS method in Fig. 2, as the discrepancies between the FBTS results and the exact results increase with increasing nonadiabaticity.  

Finally, we examine dynamics of the phonon degrees of freedom in the Holstein model. As discussed above, the CFBT and the MTEF methods can be exact in the nonadiabatic
limit, $t_0/\omega_0 \ll 1$. On the other hand, 
most methods that are accurate in the nonadiabatic limit, including CFBT and MTEF, generally
fail to accurately describe dynamics 
in the adiabatic regime. This is due to the accumulation of electron-phonon correlation during the relaxation process,
as seen from Fig. \ref{fig:Ekin_holstein} (d). Hence, as a stringent test of the capabilities of the CFBT method, we investigate the phonon dynamics in the adiabatic regime.

\begin{figure}[htbp]
\centering
\includegraphics[width=0.8\columnwidth]{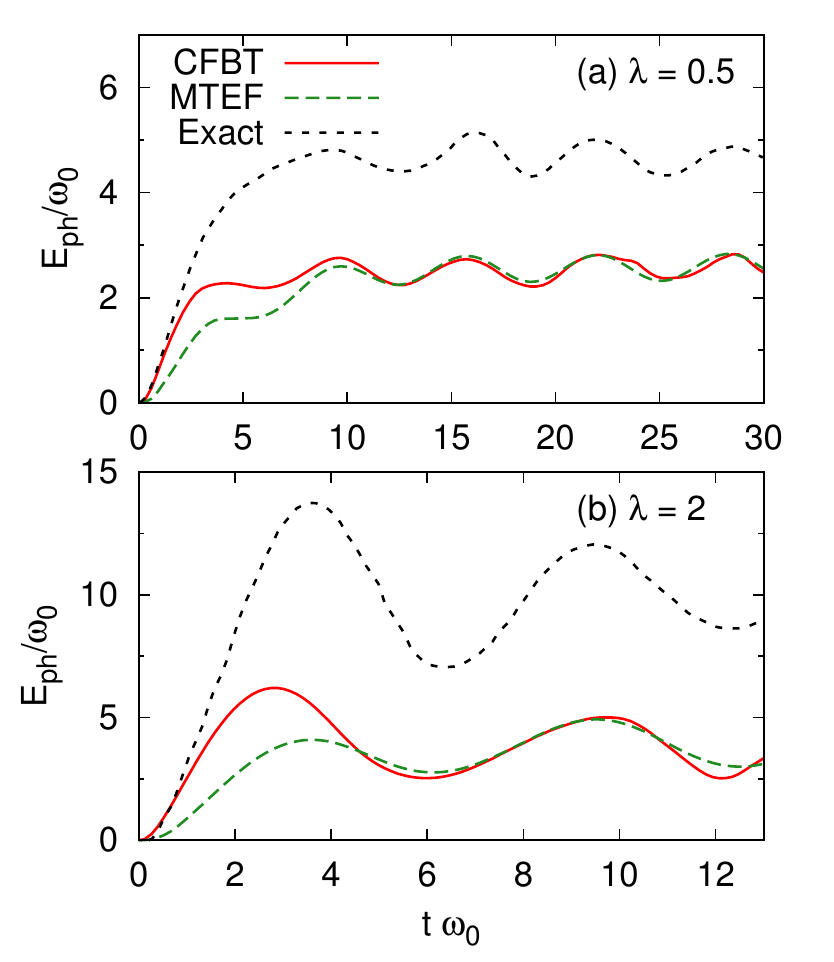}
\caption{\label{fig:Eph_holstein}
Phononic energy dynamics of the 8-site Holstein chain 
in the adiabatic regime $\omega_0=\gamma$ with (a) intermediate coupling $\lambda=0.5$ 
and (b) strong coupling $\lambda=2$.
The numerically exact solution (exact) is taken from Ref. \cite{PhysRevB.91.104302}. 
Line styles are as given in Fig. 1.}
\end{figure}

Figure \ref{fig:Eph_holstein} shows the phonon energy dynamics,
$E_{ph}(t)=\left \langle \hat H_{ph}(t) \right \rangle$, 
in the adiabatic regime $\omega_0 = \gamma$ with intermediate ($\lambda= 0.5$) 
and strong ($\lambda=2$) coupling. Here, we employed a 8-site chain instead of the 12-site chain used previously. As a proper expression of pure bath operators in the FBTS method is not trivial, 
we omit comparisons with the FBTS method for this property here. As seen from Fig. \ref{fig:Eph_holstein}, the CFBT method accurately
reproduces the short-time dynamics of the exact solution in both intermediate (a)
and strong (b) coupling regimes. On the other hand, the MTEF method fails to reproduce
the short time behavior of the exact solution.
This fact indicates that the extension of the single-product ansatz in the CFBT method
enables one to capture correlations from low-order scattering process,
as only a small number of scattering can occur in the short time regime.
This finding is consistent with the finding from the short time behavior 
of the electron relaxation dynamics in Fig. \ref{fig:Ekin_holstein_short}.

As expected, both the CFBT and the MTEF methods fail to reproduce
the long-time dynamics of the exact solution in Fig. \ref{fig:Eph_holstein}
due of the accumulation of significant electron-phonon correlation.
Furthermore, formation of the Holstein polaron is indicated
in this regime \cite{PhysRevB.91.104302}.
Since the CFBT method is based on a wavefunction ansatz, one may straightforwardly
extend the ansatz by including the Holstein polaron ground state with
a variational coefficient in order to more accurately capture the correlation present in the problem. 
The extension of the ansatz based on some prior knowledge of a given system is an interesting
and important direction to consider towards realistic applications. However, it is beyond the scope of this paper.

%=========================================================================================
\section{Summary and Outlook \label{sec:summary}}
%=========================================================================================

In this work, we have introduced a simple ansatz for the wavefunction of a many-body system called the coupled forward-backward trajectory (CBFT) method. We arrived at this ansatz by considering the phase-averaged expression for a time-dependent observable (in Eq. (\ref{eq:td-observable-phase-average})), for a wavefunction contains a pair of forward and backward propagating trajectories. We then derived the coupled equations of motion for these forward-backward trajectories through the variational principle.

We examined the properties of the CBFT method in the single-mode spin-boson model and the Holstein model. For the single-mode spin-boson model, the CFBT method shows substantial improvement in accuracy compared with the other available semi-classical trajectory based methods. The CFBT method accurately reproduces the recurrence in the population dynamics particularly well in the weak coupling regime, whereas all other approximate methods investigated fail. This indicates that the CFBT method correctly captures the quantum coherence of the bath dynamics. For the Holstein model, the CFBT method provides highly accurate electronic dynamics in the nonadiabatic regime, $\omega_0/t_0 \gg 1$. This is a somewhat surprising result since the CFBT method is based on the semiclassical trajectories. We conclude that the coupling between the forward and backward propagating semi-classical trajectories are significant to capture the quantum coherence of the nuclear wave-packet dynamics. 

On the other hand, the CFBT encounters some difficulty in the adiabatic regime, where numerous electron-phonon collisions occur, and substantial electron-phonon correlation can develop as the system approaches equilibrium. However, this is a generally challenging aspect of the theoretical description of relaxation dynamics in the adiabatic regime that any method must face. Quite encouragingly, we found that the CFBT method still provides very accurate short-time dynamics in the adiabatic regime. This improvement over mean field theory for the short-time dynamics indicates that the combination of the CFBT method and the generalized quantum master equation approach may open a new pathway to the highly efficient and accurate description relaxation dynamics in realistic correlated electron-phonon systems.

Looking toward a more complete \emph{ab-initio} simulation approach using the CFBT ansatz, the computational efficiency of the method has to be minimized in order to gain access to realistic condensed phase systems. One of our main goals in this regard is to combine the CFBT method with density functional theory. As the CFBT ansatz is perhaps the simplest possible extension of mean field theory, we expect that it can be combined with a density functional approach in a similar fashion to the development of the \emph{ab-initio} molecular dynamics approach \cite{PhysRevLett.98.023001,J.Chem.Phys.130.124107,J.Chem.Phys.127.064103,J.Chem.Phys.128.154111}. To further improve the numerical efficiency of the CFBT calculations, two central issues need to be addressed. First, efficient time-propagation algorithms for the CFBT equations of motion need to be developed. The challenge in this respect lies in the fact that the equations of motion for the electronic and ionic systems are nontrivially coupled; the structure of the electronic propagator is not Schr\"odinger-like, nor do the ions evolve classically. Nevertheless, we expect that adaptive integration schemes (not explored in the present work) may be well suited to this task. Second, and likely more importantly, the Monte Carlo sampling procedure for the bath phase space needs to be optimized. As discussed in Sec. \ref{sec:method}, the CFBT simulations using the correlated Gaussian distribution of Eq. (\ref{eq:correlated-gauss}) require more sampling effort compared to MTEF. This is due to the complex phase factor that comes from inner products of different coherent states, $\langle z|z'\rangle$, in the integrand of the expression for an observable. As the computational cost is proportional to the number of trajectories, the reduction of the trajectory number is critical to reduce the total computational cost. Work along these lines, as well as the combination of the CFBT method with the GQME approach, is already underway.

%=========================================================================================
\section*{Acknowledgements}
%=========================================================================================

We acknowledge financial support from the European Research Council(ERC-2015-AdG-694097), Grupos Consolidados (IT578-13), European Union's H2020  programme under GA no.676580 (NOMAD) and Alexander von Humboldt Foundation. AK acknowledges funding from the National Sciences and Engineering Research Council of Canada (NSERC) Discovery grant program, and start-up funding from Dalhousie University. 

\appendix 
%=========================================================================================
\section{Detailed derivation of equations of motion for the coupled forward-backward 
trajectory method
\label{appendix:deriveation}}
%=========================================================================================

Here, we describe the detailed drivation of the equations of motion for
the coupled forward-backward trajectory method [ 
Eq. (\ref{eq:eom-sub}) and Eq. (\ref{eq:eom-bath})].

First, we rewrite the Lagrangian of Eq. (\ref{eq:lagrangian00}) with
the ansatz wavefunction of Eq (\ref{eq:ansatz-wf}) as
\be
L = \frac{i\hbar}{2} \bigg [&& 
\langle \alpha | \dot \alpha \rangle  
+ \langle \alpha | \alpha \rangle  \langle z| \dot z \rangle
+ \langle \alpha | \dot \beta \rangle \langle z| z' \rangle
+ \langle \alpha | \beta \rangle \langle z| \dot z' \rangle \nonumber \\ 
&&+ \langle \beta | \dot \alpha \rangle \langle z'|z\rangle
+ \langle \beta | \alpha \rangle \langle z'| \dot z\rangle
+ \langle \beta | \dot \beta \rangle
+ \langle \beta | \beta \rangle \langle z'|\dot z'\rangle \nonumber \\
&&- \langle \dot \alpha | \alpha \rangle
-  \langle \alpha | \alpha \rangle \langle \dot z|z\rangle
- \langle \dot \alpha| \beta \rangle \langle z | z'\rangle
- \langle \alpha | \beta \rangle \langle \dot z  | z' \rangle \nonumber \\
&&- \langle \dot \beta | \alpha \rangle \langle z'|z\rangle
- \langle \beta|\alpha\rangle \langle \dot z'|z\rangle
- \langle \dot \beta | \beta \rangle 
- \langle \beta | \beta \rangle \langle \dot z'| z'\rangle 
\bigg] \nonumber \\
&& - \langle \tilde \psi| \hat H | \tilde \psi \rangle.
\label{eq:lagrangian01}
\ee

In order to evaluate the Euler-Lagrange equation of Eq. (\ref{eq:euler-lagrange-sub}),
we calculate derivatives of the Lagrangian as follows:
\be
\frac{\partial L}{\partial \langle \alpha |} = \frac{i\hbar}{2} \bigg[&&
|\dot \alpha \rangle + |\alpha \rangle \langle z|\dot z\rangle
+ | \dot \beta \rangle \langle z| z'\rangle 
+ |  \beta \rangle \langle z|\dot z'\rangle  \nonumber \\
&&- |\alpha\rangle \langle \dot z | z \rangle
-|\beta \rangle \langle \dot z | z'\rangle \bigg ] \nonumber \\
&&- \langle z| \hat H | z \rangle  |\alpha \rangle
- \langle z| \hat H | z' \rangle |\beta \rangle,
\ee
\be
\frac{\partial L}{\partial \langle \dot \alpha | } =
\frac{i\hbar}{2} \bigg [
-|\alpha \rangle - | \beta \rangle \langle z | z' \rangle 
\bigg ],
\ee
and hence,
\be
\frac{d}{dt}\frac{\partial L}{\partial \langle \dot \alpha | } =
-\frac{i\hbar}{2} \bigg [
|\dot \alpha \rangle + | \dot \beta \rangle \langle z|z'\rangle
+ | \beta \rangle \langle \dot z | z' \rangle 
+ | \beta \rangle \langle z | \dot z'\rangle
\bigg ]. \nonumber \\
\ee

By inserting these expressions into the Euler-Lagrange equation of 
Eq. (\ref{eq:euler-lagrange-sub}), the equation of motion for the subsystem
[Eq. (\ref{eq:eom-sub})] can be obtained.

Then, we describe the detailed derivation of the equation of motion of baths
[Eq. (\ref{eq:eom-bath})]. For this purpose, we start from this explicit expression
of coherent states:
\be
|z\rangle = e^{-\frac{|z|^2}{2}}e^{z \hat a ^{\dagger}}|0\rangle,
\ee
where $|0\rangle$ is the ground state of the harmonic oscillator, and 
$z\hat a^{\dagger}$ denotes $\sum_n z_n\hat a^{\dagger}_n$.
Using the explicit expression, the following expressions are obtained:
\be
\langle z'|z\rangle &=& \exp \left [ 
-\frac{|z|^2}{2}-\frac{|z'|^2}{2}+z'^*z
\right ] \nonumber \\
&=&
\exp \left [ 
-\frac{1}{2} | z-z'|^2 + i \Im
\left [zz'^* \right]
\right ],
\label{eq:ovlp-coherent-state}
\ee
and
\be
\langle z'|\dot z\rangle &=& \langle z' |
\left[ -\frac{z^* \dot z + \dot z^*z}{2} + \dot z \hat a^{\dagger}\right] |z\rangle 
\nonumber \\
&=&
\left[ -\frac{z^* \dot z + \dot z^*z}{2} + \dot z z'^* \right] 
\langle z' |z\rangle.
\label{eq:ovlp-coherent-state-dot}
\ee

Inserting these expressions (and the associated complex conjugates) into 
Eq. (\ref{eq:lagrangian01}), the Lagrangian can be explicitly described
by $z$, $\dot z$, (and their complex conjugates). Therefore, one can easily evaluate the Euler-Lagrange equation for the bath system, 
Eq. (\ref{eq:euler-lagrange-bath}).
In order to derive the equation of motion from the Euler-Lagrange equation, here we evaluate derivatives of the Lagrangian separately as follows:
\be
\frac{\partial L}{\partial z^*_n} = \frac{i\hbar}{2}
\bigg [&&
\dot z_n \langle \alpha | \alpha \rangle
+ \frac{\dot z_n}{2} \left (
\langle \alpha|\beta \rangle \langle z|z'\rangle
-\langle \beta|\alpha \rangle \langle z'|z\rangle
\right ) \nonumber \\
&&+\dot z'_n \langle \alpha|\beta\rangle \langle z|z'\rangle \nonumber \\
&&
+\left(-\frac{z_n}{2} + z'_n \right )
\left ( \langle \alpha, z| \vec{\frac{d}{dt}} |\beta,z'\rangle 
-\langle \alpha, z| {\frac{\cvec d}{dt}} |\beta,z'\rangle  \right ) \nonumber \\
&&-\frac{z_n}{2} \left (
\langle \beta,z'|\vec{\frac{d}{dt}} | \alpha, z\rangle 
-\langle \beta,z'| {\frac{\cvec d}{dt}} | \alpha, z\rangle 
\right )
\bigg ] \nonumber \\
&& - \frac{\partial}{\partial z^*_n} \langle \psi | \hat H | \psi \rangle,
\ee
where $\langle \alpha,z| \cvec{\frac{d}{dt}}$ is defined as
$\langle \dot \alpha|\otimes\langle z| + \langle \alpha|\otimes\langle \dot z|$.
One may also have
\be
\frac{\partial L}{\partial \dot z^*_n} = \frac{i\hbar}{2} \bigg[&&
-z_n \langle \alpha| \alpha \rangle -z'_n \langle \alpha|\beta\rangle\langle z|z'\rangle
\nonumber \\
&&+\frac{z_n}{2}\left (
\langle \alpha|\beta \rangle \langle z|z'\rangle
-\langle \beta|\alpha \rangle \langle z'|z\rangle
\right )
\bigg ],
\ee
and hence
\be
\frac{d}{dt}\frac{\partial L}{\partial \dot z^*} = \frac{i\hbar}{2}\bigg [ &&
-\dot z_n \langle \alpha|\alpha \rangle - z_n \frac{d}{dt} \langle \alpha|\alpha \rangle
-\dot z'_n \langle \alpha|\beta \rangle \langle z|z'\rangle
\nonumber \\
&&+\frac{\dot z_n}{2}\left( 
\langle \alpha|\beta \rangle \langle z|z'\rangle
-\langle \beta|\alpha \rangle \langle z'|z\rangle
\right )
\nonumber \\
&&+\frac{z_n}{2} \left (
\langle \alpha,z| \frac{\cvec d}{dt} | \beta,z'\rangle
+\langle \alpha,z| \vec {\frac{d}{dt}} | \beta,z'\rangle \right ) \nonumber \\
&&-\frac{z_n}{2} \left (
\langle \beta,z'| \frac{\cvec d}{dt} | \alpha,z\rangle
+\langle \beta,z'| \vec{\frac{d}{dt}} | \alpha,z\rangle
\right ) \nonumber \\
&&-z'_n \left (
\langle \alpha,z| \frac{\cvec d}{dt} | \beta,z'\rangle
+\langle \alpha,z| \vec {\frac{d}{dt}} | \beta,z'\rangle 
\right )
\bigg ]. \nonumber \\
\ee

Inserting these expressions into the Euler-Lagrange equation, 
the equation of motion for the bath [Eq. (\ref{eq:eom-bath})] can be obtained.

For harmonic baths in the case of bilinear system-bath coupling, 
the right hand side of Eq. (\ref{eq:eom-bath}) can be simply described as
Eq. (\ref{eq:force_for_cs}). Here, we describe its detailed derivation.
The Hamiltonian of such systems can be written as
\be
\hat H = \hat H_s - \gamma \sum_n \left ( 
\hat a^{\dagger}_n + \hat a_n
\right ) \otimes \hat \Gamma_n 
+ \sum_n \hbar \omega_n \left ( 
\hat a^{\dagger}_n \hat a_n + \frac{1}{2}
\right ), \nonumber \\
\ee
where $\hat \Gamma_n$ is a pure subsystem operator.
Then the expectation value of the Hamiltonian with the ansatz wavefunction
of Eq. (\ref{eq:ansatz-wf}) can be expressed as
\be
&&\langle \tilde \psi | \hat H | \tilde \psi \rangle =
\langle \alpha |\hat H_s | \alpha \rangle
+\langle \beta |\hat H_s | \beta \rangle \nonumber \\
&&
+ \langle \alpha |\hat H_s | \beta \rangle \langle z|z'\rangle 
+ \langle \beta |\hat H_s | \alpha \rangle \langle z'|z\rangle \nonumber \\
&&+ \sum_n \bigg [ 
\hbar \omega_n \left( z^*_n z_n + \frac{1}{2} \right) \langle \alpha|\alpha \rangle
+\hbar \omega_n \left( z'^*_n z'_n + \frac{1}{2} \right) 
\langle \beta|\beta \rangle \nonumber \\
&&+\hbar \omega_n \left( z^*_nz'_n + \frac{1}{2} \right) \langle \alpha | \beta \rangle
\langle z| z'\rangle \nonumber \\
&&+\hbar \omega_n \left( z'^*_nz_n + \frac{1}{2} \right) \langle \beta | \alpha \rangle
\langle z'| z\rangle \nonumber \\
&&-\gamma (z^*_n+z_n) \langle \alpha|\hat \Gamma_n | \alpha \rangle
-\gamma (z'^*_n+z'_n) \langle \beta|\hat \Gamma_n | \beta \rangle \nonumber \\
&&-\gamma \left( z^*_n + z'_n \right) \langle \alpha|\hat \Gamma_n|\beta \rangle 
\langle z|z'\rangle
\nonumber \\
&&-\gamma \left( z'^*_n + z_n \right)\langle \beta|\hat \Gamma_n|\alpha \rangle 
\langle z'|z\rangle
\bigg ]
\ee

Using the expression of Eq. (\ref{eq:ovlp-coherent-state}),
one can explicitly evaluate the derivative of 
$\langle \tilde \psi|\hat H|\tilde \psi\rangle$ by $z^*_n$ and 
obtain Eq. (\ref{eq:force_for_cs}).

%=========================================================================================
\section{Reduction to Ehrenfest dynamics  \label{appendix:mtef}}
%=========================================================================================

Here, we derive the multi-trajectory Ehrenfest dynamics (MTEF) method based on the coherent state expansion of Eq. (\ref{eq:td-observable}), and clarify a relation between 
MTEF and the coupled forward-backward trajectory (CFBT) method.
First, we assume that the initial density matrix $\hat \rho$ is not entangled, and thus, can be described as \cite{PhysRevA.40.4277},
\be
\hat \rho = \sum_n w_n \hat \rho_{s,n} \otimes \hat \rho_{b,n},
\label{eq:unentangled-dm}
\ee
with probability $w_n$, where $\hat \rho_{s,n}$ and $\hat \rho_{b,n}$ are density matrices of subsystem and bath, respectively.
Furthermore, the subsystem density matrix $\hat \rho_{s,n}$ can be decomposed as
\be
\hat \rho_{s,n} = \sum_i \lambda_{n,i} | \alpha_{n,i} \rangle \langle \alpha_{n,i}|,
\ee
where $|\alpha_{n,i}\rangle$ and $\lambda_{n,i}$ are eigenstate and eigenvalue of 
$\hat \rho_{s,n}$, respectively.
Therefore, unentangled density matrices can be described by a linear combination of
direct products $|\alpha\rangle\langle\alpha|\otimes \hat \rho_b$.

Hereafter, for simplicity, we assume that the total unentangled density matrix $\hat \rho$ can be 
factorized as $\hat \rho = |\alpha \rangle \langle \alpha| \otimes \hat \rho_b$.
However, as descussed above, this assumption does not loose generality of the discussion.
Under this assumption, the observable in Eq. (\ref{eq:td-observable}) can be rewritten as
\be
\left \langle \hat B(t) \right \rangle &=& 
\int \frac{d^2z}{\pi^{N_b}} \int \frac{d^2z'}{\pi^{N_b}}
|\langle z| z' \rangle |^2
\frac{\langle z'| \hat \rho_b |z \rangle }{\langle z'| z \rangle }
\nonumber \\
&& \times
\frac{
\left \{ \langle \alpha | \otimes \langle z|  \right \}
\hat B(t)
\left \{  |\alpha \rangle \otimes |z' \rangle \right \}
}{\langle z| z' \rangle }.
\label{eq:appendix-observable}
\ee

From Eq. (\ref{eq:ovlp-coherent-state}), it is expected that the overlap of two coherent states decays rapidly in the phase space. 
On top of this fact, we assume that the overlap $\langle z|z'\rangle$ decays sufficiently 
rapidly so that the rest of the integrand, 
$\langle z'|\hat \rho_b |z\rangle 
\left \{ \langle \alpha | \otimes \langle z|  \right \}
\hat B(t)
\left \{  |\alpha \rangle \otimes |z' \rangle \right \}
/|\langle z| z' \rangle|^2
$, 
can be evaluated by imposing 
$z=z'$.

This assumption can be realized by treating the overlap of the coherent states 
as a delta function;
\be
|\langle z| z'\rangle|^2 = \pi^{N_b}\delta(z-z').
\label{eq:delta-overlap}
\ee 
%The same assumption is employed in 
%the derivation of the forward-backward trajectory solution approach 
%\cite{JChemPhys137.22A507,JChemPhys138.134110}. 
Based on this assumption, one can approximate the observable as
\be
\left \langle \hat B(t) \right \rangle &\simeq& 
\int \frac{d^2z}{\pi^{N_b}} \int \frac{d^2z'}{\pi^{N_b}}
\pi^{N_b} \delta(z-z')
\frac{\langle z'| \hat \rho_b |z \rangle }
{\langle z'|z\rangle}
\nonumber \\
&& \times
\frac{
\left \{ \langle \alpha | \otimes \langle z|  \right \}
\hat B(t)
\left \{  |\alpha \rangle \otimes |z' \rangle \right \}
}{\langle z| z' \rangle } \nonumber \\
&=& 
\int \frac{d^2z}{\pi^{N_b}}
\langle z| \hat \rho_b |z \rangle  \nonumber \\
&&\times
\left \{ \langle \alpha | \otimes \langle z|  \right \}
\hat B(t)
\left \{  |\alpha \rangle \otimes |z \rangle \right \}.
\label{eq:observable-EF}
\ee

Then, we approximate the time-propagation of the wavefunction; 
$|\psi \rangle = \hat U(0,t)\left \{  |\alpha \rangle \otimes |z \rangle \right \}$.
For this purpose, we introduce the following ansatz;
\be
|\psi_{EF} (t)\rangle = |\alpha (t) \rangle \otimes | z(t) \rangle,
\label{eq:EF-ansatz}
\ee
where the total wavefunction is the direct product of the subsystem state and the bath coherent state at all the times.

To derive the equation of motion for the ansatz wavefunction, $|\psi_{EF}(t)\rangle$,
we consider the following Lagrangian;
\be
L &=& i \frac{
\langle \psi_{EF}(t)| \dot \psi_{EF}(t)\rangle
-\langle \dot \psi_{EF}(t)| \psi_{EF}(t)\rangle
}{2} \nonumber \\
&& -\langle \psi_{EF}(t)| \hat H | \psi_{EF}(t)\rangle.
\label{eq:lag-ef}
\ee

Under the orthogonal approximation for the coherent states
one has the following relation,
\be
\langle z| a_i a^{\dagger}_j | z \rangle 
&=& \langle z| a_i \int \frac{d^2z'}{\pi^{N_b}}
|z'\rangle \langle z'|
a^{\dagger}_j | z \rangle \nonumber \\
&=& \int \frac{d^2z'}{\pi^{N_b}}  z'_iz'^*_j \langle z|z'\rangle 
\langle z'| z \rangle \nonumber \\
&\simeq& 
\int \frac{d^2z'}{\pi^{N_b}}  z'_iz'^*_j \pi^{N_b}\delta(z-z')
 \nonumber \\
&=&  z_iz^*_j.
\ee
Similar approximations can be construct for higher order combinations of $\hat a$ and $\hat a^{\dagger}$.

Using these approximation, all the creation and annihilation operators, $a$ and $a^{\dagger}$,
in the Lagrangian (\ref{eq:lag-ef}) 
can be approximated by $c$-numbers, $z$ and $z^*$.
Then, the Lagrangian can be approximated as
\be
L &\simeq& i \frac{
\langle \psi_{EF}(t)| \dot \psi_{EF}(t)\rangle
-\langle \dot \psi_{EF}(t)| \psi_{EF}(t)\rangle
}{2} \nonumber \\
&& -\langle \alpha(t)| \hat H_{EF}(z,z^*) | \alpha(t)\rangle,
\label{eq:lag-ef-approx}
\ee
where the effective Hamiltonian $\hat H_{EF}(z,z^*)$ is defined by replacing
$a$ and $a^{\dagger}$ by $c$-numbers, $z$ and $z^*$, respectively.

Inserting the ansatz of Eq. (\ref{eq:EF-ansatz}) into Eq. (\ref{eq:lag-ef-approx})
and using Eq. (\ref{eq:ovlp-coherent-state-dot}),
the Lagrangian can be rewritten as
\be
L &=& i \hbar \frac{ \langle \alpha | \dot \alpha\rangle 
- \langle \dot \alpha | \alpha \rangle}{2}
- \langle \alpha | \hat H_{EF}(z,z^*)|\alpha \rangle
+\Re \left [i\hbar z^* \dot z \right ] \nonumber \\
&=&
i \hbar \frac{ \langle \alpha | \dot \alpha\rangle 
- \langle \dot \alpha | \alpha \rangle}{2}
- \langle \alpha | \hat H_{EF}(q,p)|\alpha\rangle 
+\frac{1}{2}\left (p\dot q - q \dot p \right ).
\nonumber \\
\label{eq:lag-ef-approx2}
\ee
In the last line, the canonical transformation,
$z~=~\sqrt{\frac{m\omega}{2\hbar}} \left (q+\frac{i}{m\omega}p \right )$,
is applied. Then, one may derive the equations of motion for $q(t)$, $p(t)$ and $|\alpha(t)\rangle$ based on the Euler-Lagrange equation. These equations of motion are nothing but those of the Ehrenfest dynamics; 
\be
&& i\hbar \frac{\partial}{\partial t}| \alpha(t)\rangle = 
\hat H_{EF}(q(t),p(p)) | \alpha(t)\rangle, \label{eq:EF-psi} \\
&&\frac{d}{dt} q_j(t) = 
\frac{\partial}{\partial p_j} \langle \alpha(t) | \hat H_{EF} (q,p) | \alpha (t) \rangle 
= \frac{p_j(t)}{m}, \label{eq:EF-q}   \\
&& \frac{d}{dt} p_j(t) = - \frac{\partial}{\partial q_j} \langle \alpha(t) 
| \hat H_{EF} (q,p) | \alpha (t) \rangle
 = - \frac{\partial V(q)}{\partial q_j}.  \nonumber \\
\label{eq:EF-p}
\ee
For the right hand sides of Eq. (\ref{eq:EF-q}) and  Eq. (\ref{eq:EF-p}), 
we assumed the conventional form of the Hamiltonian;
$\langle \alpha| \hat H_{EF} | \alpha \rangle = p^2/2m + V(q)$.

By using the solutions of these equations, Eq. (\ref{eq:observable-EF}) can be easily evaluated. Since contributions from multi trajectories are simply summed in Eq. (\ref{eq:EF-p}) with the weight from the bath density matrix, the derived method here is nothing but the MTEF method. Therefore, we proved that the MTEF method can be derived upon the three approximations; (i) the initial density matrix is not entangled, (ii) the delta-overlap for the coherent states, Eq.(\ref{eq:delta-overlap}), and (iii) the independent trajectory ansatz, Eq. (\ref{eq:EF-ansatz}). Furthermore, since the CFBT method does not relay on Eq. (\ref{eq:unentangled-dm}) or Eq.(\ref{eq:delta-overlap}), and it employs a more general ansatz, Eq. (\ref{eq:ansatz-wf}), we can conclude that the CFBT method is an extension of the MTEF method.

%=========================================================================================
\bibliographystyle{apsrev4-1}
\bibliography{ref}

\end{document}